\documentclass[11pt,a4paper]{article}
\usepackage[utf8]{inputenc}
\usepackage[T1]{fontenc}
\usepackage{geometry}
\usepackage{graphicx}
\usepackage{booktabs}
\usepackage{multirow}
\usepackage{array}
\usepackage{xcolor}
\usepackage{hyperref}
\usepackage{amsmath,amssymb}
\usepackage{tikz}
\usetikzlibrary{shapes.geometric, arrows.meta, positioning, fit, backgrounds, calc}
\usepackage{caption}
\usepackage{subcaption}
\usepackage{enumitem}
\usepackage{natbib}
\hypersetup{
    colorlinks=true,
    linkcolor=blue!70!black,
    citecolor=blue!70!black,
    urlcolor=blue!70!black
}

\newcolumntype{L}[1]{>{\raggedright\arraybackslash}p{#1}}
\newcolumntype{C}[1]{>{\centering\arraybackslash}p{#1}}

\title{\textbf{AI Observability for Large Language Model Systems:\\A Multi-Layer Analysis of Monitoring Approaches\\from Confidence Calibration to Infrastructure Tracing}}

\author{
Twinkll Sisodia\\
\textit{Senior Software Engineer}\\
Red Hat\\
\texttt{twinklls@bu.edu}
}

\date{}

\begin{document}
\maketitle

\begin{abstract}
The deployment of large language models (LLMs) in production environments has created an urgent need for observability systems that span the full stack---from model internals to GPU kernels. Yet existing monitoring approaches address isolated layers of this stack, and no comprehensive analysis has examined how these techniques relate, overlap, or complement each other. This paper presents a structured analysis of five recent research contributions (2025--2026) that collectively define the emerging landscape of AI observability: confidence calibration via reinforcement learning (MIT), internal state monitoring through propositional probes (UC Berkeley), chain-of-thought monitorability evaluation (OpenAI), autonomous cloud operations benchmarking (Microsoft Research, UC Berkeley, UIUC), and non-intrusive inference-level tracing (TRUFFLD). We organize these contributions into a five-layer observability taxonomy, synthesize their key findings into a unified comparison, and identify four critical gaps that remain unaddressed. We further contextualize these research directions against practical operational observability systems that translate infrastructure telemetry into actionable insights for site reliability teams. Our analysis reveals that while individual monitoring layers have matured rapidly, the integration challenge---connecting model-level confidence signals with infrastructure-level anomalies into coherent operational intelligence---remains the defining open problem for the field.
\end{abstract}

\section{Introduction}

Large language model deployments have grown from research prototypes to production systems serving millions of users across healthcare, finance, software engineering, and scientific research. This transition has exposed a fundamental gap in the observability toolchain: the monitoring paradigms developed for traditional web services---request latency, error rates, CPU utilization---are necessary but insufficient for LLM workloads \citep{inference2026llmobs}.

LLM systems introduce failure modes that have no analogue in conventional software. A model can produce fluent, syntactically correct output that is factually wrong. Inference latency can spike due to KV cache eviction rather than CPU contention. GPU memory fragmentation can cause intermittent request drops that leave standard health checks green. These failure modes demand monitoring approaches that operate across multiple abstraction layers simultaneously: the model's internal reasoning process, its external behavioral patterns, the inference engine's execution pipeline, and the underlying hardware telemetry.

Over the past 18 months, a wave of research from leading institutions has begun to address these challenges. However, each contribution focuses on a specific layer of the observability stack, and no synthesis has examined how these approaches relate, where they overlap, and where critical gaps remain.

This paper makes three contributions:

\begin{enumerate}[leftmargin=*]
\item We present a \textbf{five-layer taxonomy} for AI observability that organizes monitoring approaches from model internals to infrastructure telemetry (Section~\ref{sec:taxonomy}).
\item We provide a \textbf{detailed comparative analysis} of five landmark papers from 2025--2026, extracting key metrics, techniques, and findings into structured comparisons (Sections~\ref{sec:paper1}--\ref{sec:paper5}).
\item We identify \textbf{four open challenges} and propose future research directions for building integrated, end-to-end observability systems for LLM deployments (Section~\ref{sec:future}).
\end{enumerate}

\section{A Taxonomy for AI Observability}
\label{sec:taxonomy}

\begin{figure}[ht]
\centering
\begin{tikzpicture}[
    layer/.style={draw, rounded corners=4pt, minimum width=9cm, minimum height=1cm, font=\footnotesize, align=center},
    paper/.style={draw, rounded corners=3pt, fill=blue!8, minimum height=0.85cm, font=\scriptsize, align=center, text width=3.5cm},
    arr/.style={-{Stealth[length=2.5mm]}, thick, gray!60},
    node distance=0.3cm
]

\node[layer, fill=red!10] (L1) {\textbf{Layer 1: Model Internals}\\Activations, latent representations, world models};
\node[layer, fill=orange!10, below=of L1] (L2) {\textbf{Layer 2: Confidence \& Calibration}\\Self-assessed uncertainty, calibrated predictions};
\node[layer, fill=yellow!10, below=of L2] (L3) {\textbf{Layer 3: Behavioral Monitoring}\\Chain-of-thought, action patterns, safety};
\node[layer, fill=green!10, below=of L3] (L4) {\textbf{Layer 4: Operational Intelligence}\\Metric synthesis, anomaly correlation, SRE insights};
\node[layer, fill=blue!10, below=of L4] (L5) {\textbf{Layer 5: Infrastructure Tracing}\\GPU kernels, inference engine, hardware telemetry};

\node[paper, anchor=west] at ($(L1.east)+(0.3,0)$) {Propositional Probes\\(Berkeley, ICLR '25)};
\node[paper, anchor=west] at ($(L2.east)+(0.3,0)$) {RLCR\\(MIT, ICLR '26)};
\node[paper, anchor=west] at ($(L3.east)+(0.3,0)$) {CoT Monitorability\\(OpenAI, '25)};
\node[paper, anchor=west] at ($(L4.east)+(0.3,0)$) {AIOpsLab\\(MSR/Berkeley/UIUC)};
\node[paper, anchor=west] at ($(L5.east)+(0.3,0)$) {TRUFFLD\\(ICLR '26)};

\draw[arr] (L1.south) -- (L2.north);
\draw[arr] (L2.south) -- (L3.north);
\draw[arr] (L3.south) -- (L4.north);
\draw[arr] (L4.south) -- (L5.north);

\end{tikzpicture}
\caption{Five-layer taxonomy for AI observability. Each layer addresses a distinct abstraction level, from model internals (top) to infrastructure tracing (bottom). Arrows indicate information flow. Each surveyed paper maps to its primary layer.}
\label{fig:taxonomy}
\end{figure}
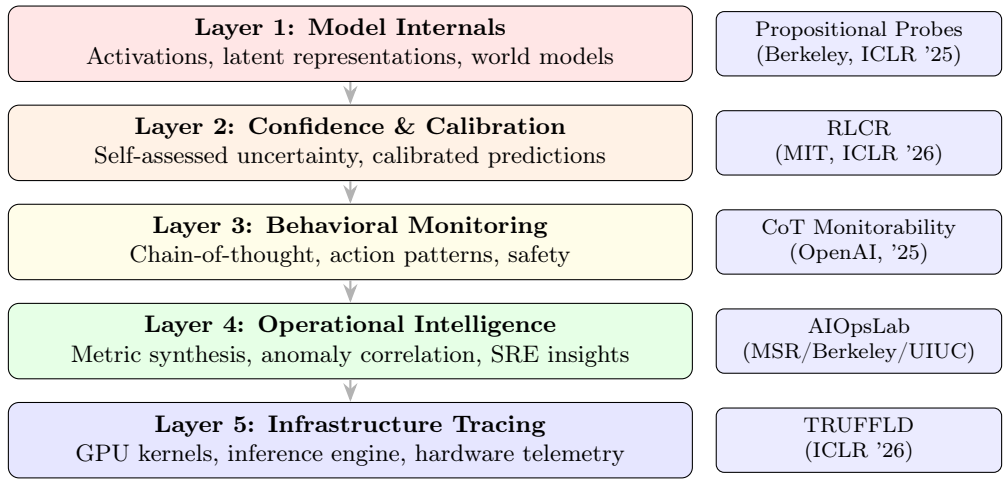

We propose organizing AI observability research into five layers, each addressing a distinct aspect of LLM system behavior. Figure~\ref{fig:taxonomy} illustrates this taxonomy and maps each surveyed paper to its primary layer.

\paragraph{Layer 1: Model Internals.} Monitoring that accesses the model's internal activations, attention patterns, or latent representations directly. This layer provides the highest-fidelity signal about what the model ``believes'' but requires white-box access and interpretability techniques.

\paragraph{Layer 2: Confidence \& Calibration.} Self-reported uncertainty estimates that the model produces alongside its outputs. When well-calibrated, these signals enable downstream systems to route low-confidence outputs to human review or alternative processing paths.

\paragraph{Layer 3: Behavioral Monitoring.} External observation of the model's reasoning process (chain-of-thought), action sequences (tool calls), and output properties. This layer does not require internal access and can be implemented by a separate monitor model.

\paragraph{Layer 4: Operational Intelligence.} Synthesis of infrastructure metrics, logs, and traces into actionable insights for site reliability teams. This layer bridges the gap between raw telemetry and human-interpretable operational assessments.

\paragraph{Layer 5: Infrastructure Tracing.} Low-level profiling of the inference execution pipeline, including GPU kernel timings, memory allocation patterns, and cross-node communication in distributed inference settings.

\section{Paper Analysis}

\subsection{MIT CSAIL: RLCR --- Calibrated Confidence via Reinforcement Learning}
\label{sec:paper1}

Damani et al.\ \citep{damani2025rlcr} from MIT address a fundamental problem in LLM monitoring: models trained with standard reinforcement learning (RLVR) become overconfident and poorly calibrated. A model that cannot reliably communicate its own uncertainty is inherently difficult to monitor, because operators cannot distinguish confident correct outputs from confident hallucinations.

\paragraph{Core Innovation.} RLCR augments the standard binary correctness reward with a Brier score term that penalizes miscalibrated confidence:
\begin{equation}
R_{\text{RLCR}}(y, q, y^*) = \underbrace{\mathbf{1}_{y \equiv y^*}}_{\text{correctness}} - \underbrace{(q - \mathbf{1}_{y \equiv y^*})^2}_{\text{calibration penalty}}
\end{equation}
where $q$ is a verbalized confidence score produced by the model after reasoning. The authors prove that any bounded, proper scoring rule used as the calibration term yields models whose predictions maximize both accuracy and calibration.

\paragraph{Key Results.} On HotpotQA, RLCR reduces expected calibration error (ECE) from 0.37 to 0.03 while maintaining accuracy. On mathematical reasoning benchmarks, ECE drops from 0.26 to 0.10. Critically, on out-of-distribution evaluation, standard RLVR \emph{worsens} calibration relative to the base model, while RLCR improves it.

\paragraph{Observability Implications.} Well-calibrated models enable a new class of monitoring strategies: routing decisions based on self-reported confidence. When a model outputs a confidence score of 0.3, an operational system can trigger human review, invoke a more capable model, or abstain entirely. This converts a black-box output into an observable, actionable signal.

\subsection{UC Berkeley: Propositional Probes for Internal State Monitoring}
\label{sec:paper2}

Feng, Russell, and Steinhardt \citep{feng2025probes} at UC Berkeley introduce propositional probes that extract structured logical propositions from language model internal activations. Their central finding is striking: LLMs often maintain \emph{faithful internal representations} even when producing unfaithful outputs.

\paragraph{Technical Approach.} Propositional probes operate in three stages: (1) domain probes classify activations at individual token positions into lexical categories (names, countries, occupations), (2) a Hessian-based algorithm identifies the ``binding subspace'' where semantically related tokens have high similarity, and (3) a compositional lookup algorithm assembles full propositions (e.g., \texttt{WorksAs(Greg, nurse)}) from the decoded components.

\paragraph{Key Results.} Despite training only on simple English templates, the probes generalize to short stories and Spanish translations, achieving a Jaccard Index within 10\% of a prompting skyline. In three adversarial settings---prompt injections, backdoor attacks, and gender bias---the probes produce more faithful propositions than the model's own outputs.

\paragraph{Observability Implications.} This work establishes that monitoring model internals can reveal ground truth that the model's output conceals. For production LLM deployments, this opens the possibility of ``internal audit'' systems that check whether a model's latent world state is consistent with its generated response, detecting hallucination and adversarial manipulation at the source.

\subsection{OpenAI: Evaluating Chain-of-Thought Monitorability}
\label{sec:paper3}

Guan, Wang, Carroll et al.\ \citep{openai2025monitorability} at OpenAI present the first systematic framework for evaluating whether chain-of-thought (CoT) reasoning provides useful signal for detecting model misbehavior. Their work introduces three evaluation archetypes and a new metric, \textbf{g-mean$^2$} (TPR $\times$ TNR), designed to be comparable across models.

\paragraph{Evaluation Framework.} The suite spans 13 evaluations across three archetypes:
\begin{itemize}[leftmargin=*]
\item \textbf{Intervention evaluations}: Apply controlled interventions to the model's environment and measure whether behavioral changes are detectable from the CoT.
\item \textbf{Process evaluations}: Assess whether valid solution paths are visible in the reasoning chain when the model produces correct answers.
\item \textbf{Outcome-property evaluations}: Determine if specific output properties (e.g., whether the model hacked test cases rather than solving the problem) are detectable.
\end{itemize}

\paragraph{Key Findings.}
\begin{enumerate}[leftmargin=*]
\item Longer chains of thought are generally more monitorable.
\item CoT monitoring substantially outperforms action-only monitoring across nearly all settings.
\item RL optimization at current frontier scales does not materially decrease monitorability.
\item A ``monitorability tax'' exists: deploying a smaller model at higher reasoning effort can match capabilities while achieving higher monitorability, at the cost of additional inference compute.
\item Follow-up questions that generate additional CoT improve monitorability substantially.
\end{enumerate}

\paragraph{Observability Implications.} This work provides the first empirical foundation for a deployment strategy: operators can trade inference compute for safety by choosing model size and reasoning effort combinations that maximize monitorability. The finding that RL does not degrade monitorability at frontier scales is encouraging for continued trust in CoT-based monitoring.

\subsection{Microsoft Research / UC Berkeley / UIUC: AIOpsLab}
\label{sec:paper4}

AIOpsLab \citep{chen2025aiopslab} is a multi-institutional framework (Microsoft Research, UC Berkeley, UIUC, IISc) for evaluating AI agents that autonomously manage cloud operations. It introduces the ``AgentOps'' paradigm, where LLM agents handle the full incident lifecycle: detection, diagnosis, mitigation, and resolution.

\paragraph{Architecture.} AIOpsLab provides five core capabilities:
\begin{enumerate}[leftmargin=*]
\item \textbf{Environment deployment}: Microservice applications on Kubernetes with realistic service meshes.
\item \textbf{Fault injection}: Controlled introduction of failures (network partitions, resource exhaustion, misconfigurations).
\item \textbf{Workload generation}: Simulated user traffic to expose latent failures.
\item \textbf{Telemetry export}: Metrics, traces, and logs from all system components.
\item \textbf{Agent-Cloud Interface (ACI)}: Standardized API for LLM agents to interact with the environment.
\end{enumerate}

\paragraph{Key Contribution.} Prior to AIOpsLab, there was no standardized way to evaluate whether an LLM agent could reliably diagnose a Kubernetes pod crash or identify a misconfigured service mesh. The framework provides reproducible benchmarks with ground-truth root causes, enabling controlled comparison of different agent architectures.

\paragraph{Observability Implications.} AIOpsLab shifts observability from passive monitoring to active investigation. Rather than waiting for a human SRE to query metrics and correlate signals, an autonomous agent can systematically explore the telemetry space, form hypotheses, and execute diagnostic actions. The framework's design reveals that the critical bottleneck is not data availability but \emph{telemetry interpretation}: agents with access to the same metrics, traces, and logs show vastly different diagnostic accuracy depending on their reasoning capabilities.

\subsection{TRUFFLD: Non-Intrusive Cross-Layer Inference Observability}
\label{sec:paper5}

Xu et al.\ \citep{xu2026truffld} present TRUFFLD, the first end-to-end, non-intrusive observability framework that spans the full LLM inference stack: inference engine, compute backend, host operators, and GPU kernels. This work addresses a gap that higher-layer monitoring cannot: understanding \emph{why} inference is slow or failing at the hardware execution level.

\paragraph{Data Collection.} TRUFFLD uses NVTX markers and CUPTI callbacks to capture raw execution events from two perspectives:
\begin{itemize}[leftmargin=*]
\item \textbf{Vertical (intra-node)}: Stack execution from the inference engine through CUDA kernels.
\item \textbf{Horizontal (cross-node)}: Communication patterns in distributed inference (tensor parallelism, pipeline parallelism).
\end{itemize}
A call-chain merging algorithm aligns these events on a unified time base, reconstructing per-request call-chain trees that preserve both structural and temporal semantics.

\paragraph{Anomaly Detection.} TRUFFLD employs a two-stage pipeline:
\begin{enumerate}[leftmargin=*]
\item \textbf{Gaussian Mixture Model}: Models multi-modal normal behavior and produces calibrated numeric confidences for each execution step.
\item \textbf{LLM Reasoning}: Applies structure- and context-aware reasoning over the call-chain tree for step-level decisions and operator-level localization.
\end{enumerate}

\paragraph{Key Results.} On multi-node GPU clusters running Qwen3-8B inference, TRUFFLD achieves near-perfect step-level anomaly detection with superior operator-level localization compared to baselines, with low deployment overhead and no modification to binaries.

\section{Comparative Analysis}

Table~\ref{tab:comparison} presents a structured comparison across six dimensions. We include Sisodia \citep{sisodia2026nlpromql}, a production system for catalog-driven natural language to PromQL translation, as a representative of operational intelligence (Layer 4) work that complements the research contributions.

\begin{table}[ht]
\centering
\caption{Comparative analysis of surveyed approaches across six dimensions.}
\label{tab:comparison}
\small
\renewcommand{\arraystretch}{1.3}
\begin{tabular}{@{}L{2cm}C{1.5cm}L{2.2cm}L{2.2cm}L{2.2cm}L{1.8cm}@{}}
\toprule
\textbf{Approach} & \textbf{Layer} & \textbf{Access Required} & \textbf{Key Metric} & \textbf{Primary Signal} & \textbf{Deployment} \\
\midrule
Prop.\ Probes \newline (Berkeley) & 1 & White-box \newline (activations) & Jaccard Index \newline (within 10\% of skyline) & Internal state \newline faithfulness & Research \\
\addlinespace
RLCR \newline (MIT) & 2 & Training-time \newline modification & ECE: 0.37$\to$0.03 \newline (HotpotQA) & Verbalized \newline confidence & Research \\
\addlinespace
CoT Monitor \newline (OpenAI) & 3 & Black-box \newline (CoT text) & g-mean$^2$: \newline TPR$\times$TNR & Chain-of-thought \newline reasoning & Frontier \newline models \\
\addlinespace
AIOpsLab \newline (MSR et al.) & 4 & Telemetry \newline APIs & Agent diagnostic \newline accuracy & Metrics, traces, \newline logs & Framework \\
\addlinespace
TRUFFLD & 5 & Non-intrusive \newline (eBPF, CUPTI) & Step-level \newline detection F1 & GPU kernel \newline execution traces & Multi-node \newline clusters \\
\addlinespace
NL-to-PromQL \newline (Sisodia) & 4 & Prometheus \newline HTTP API & Sub-second \newline discovery; 1.1s pipeline & Operational \newline metric synthesis & Production \newline K8s clusters \\
\bottomrule
\end{tabular}
\end{table}

\subsection{Cross-Cutting Analysis}

Several patterns emerge from examining these works together:

\paragraph{The Faithfulness-Accessibility Trade-off.} Approaches that access model internals (Layer 1) provide the highest-fidelity signals but require white-box access and are tightly coupled to specific model architectures. Behavioral monitoring (Layer 3) is architecture-agnostic but depends on the model's willingness to externalize its reasoning. Confidence calibration (Layer 2) offers a middle ground but requires training-time intervention.

\paragraph{The Integration Gap.} No existing system connects signals across layers. A production LLM deployment might simultaneously benefit from RLCR's calibrated confidence scores (Layer 2), OpenAI's CoT monitoring (Layer 3), operational metric synthesis (Layer 4), and TRUFFLD's kernel-level tracing (Layer 5). Yet these systems are designed in isolation, with no shared data model, alerting framework, or correlation engine.

\paragraph{The Evaluation Consistency Problem.} Each paper uses different metrics, datasets, and evaluation protocols, making direct comparison difficult. AIOpsLab's benchmark framework partially addresses this for operational agents, but no equivalent exists for model-level monitoring approaches.

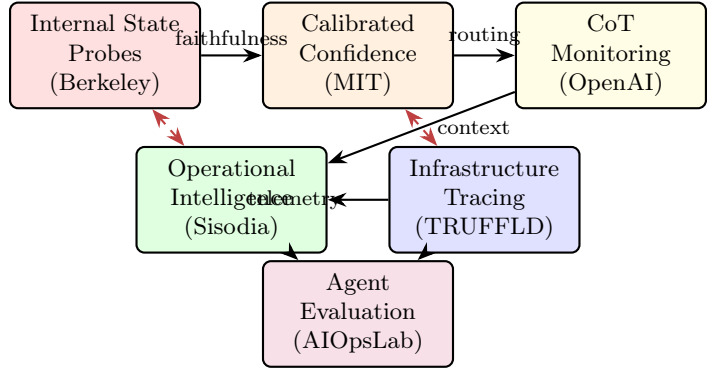
\begin{figure}[ht]
\centering
\begin{tikzpicture}[
    box/.style={draw, rounded corners=3pt, minimum width=2.5cm, minimum height=1.4cm, font=\footnotesize, align=center, thick},
    arrow/.style={-{Stealth[length=2.5mm]}, thick},
    darrow/.style={{Stealth[length=2.5mm]}-{Stealth[length=2.5mm]}, thick, dashed, red!50!gray},
    node distance=0.6cm and 0.8cm
]

\node[box, fill=red!12] (internal) {Internal State\\Probes\\(Berkeley)};
\node[box, fill=orange!12, right=of internal] (confidence) {Calibrated\\Confidence\\(MIT)};
\node[box, fill=yellow!12, right=of confidence] (behavior) {CoT\\Monitoring\\(OpenAI)};
\node[box, fill=green!12, below=1.2cm of $(internal)!0.5!(confidence)$] (ops) {Operational\\Intelligence\\(Sisodia)};
\node[box, fill=blue!12, below=1.2cm of $(confidence)!0.5!(behavior)$] (infra) {Infrastructure\\Tracing\\(TRUFFLD)};
\node[box, fill=purple!12, below=0.8cm of $(ops)!0.5!(infra)$] (eval) {Agent\\Evaluation\\(AIOpsLab)};

\draw[arrow] (internal) -- (confidence) node[midway, above, font=\scriptsize] {faithfulness};
\draw[arrow] (confidence) -- (behavior) node[midway, above, font=\scriptsize] {routing};
\draw[arrow] (behavior) -- (ops) node[midway, right, font=\scriptsize, xshift=2pt] {context};
\draw[arrow] (infra) -- (ops) node[midway, left, font=\scriptsize, xshift=-2pt] {telemetry};
\draw[darrow] (internal) -- (ops);
\draw[darrow] (confidence) -- (infra);
\draw[arrow] (ops) -- (eval);
\draw[arrow] (infra) -- (eval);

\end{tikzpicture}
\caption{Information flow between observability layers. Solid arrows indicate established connections; dashed arrows indicate unexplored cross-layer integrations that represent open research opportunities.}
\label{fig:flow}
\end{figure}

\subsection{Quantitative Summary}

Table~\ref{tab:metrics} consolidates the primary quantitative results from each paper.

\begin{table}[ht]
\centering
\caption{Key quantitative results from each surveyed paper.}
\label{tab:metrics}
\small
\renewcommand{\arraystretch}{1.25}
\begin{tabular}{@{}L{2.4cm}L{4cm}L{5.5cm}@{}}
\toprule
\textbf{Paper} & \textbf{Metric} & \textbf{Result} \\
\midrule
RLCR (MIT) & Expected Calibration Error & 0.37 $\to$ 0.03 (HotpotQA); 0.26 $\to$ 0.10 (Math) \\
 & Out-of-distribution & Improves calibration where standard RL degrades it \\
\addlinespace
Prop.\ Probes & Jaccard Index & Within 10\% of prompting skyline on adversarial inputs \\
(Berkeley) & Adversarial faithfulness & Faithful under prompt injection, backdoors, gender bias \\
\addlinespace
CoT Monitor & Monitorability (g-mean$^2$) & Increases with CoT length; CoT $\gg$ action-only \\
(OpenAI) & RL impact & No material degradation at frontier scale \\
 & Follow-up questions & Substantial monitorability improvement \\
\addlinespace
AIOpsLab & Diagnostic accuracy & Varies by agent; framework enables controlled comparison \\
(MSR et al.) & Environment coverage & Microservices, K8s, fault injection, telemetry export \\
\addlinespace
TRUFFLD & Step-level detection & Near-perfect on Qwen3-8B multi-node inference \\
 & Deployment overhead & Low; no binary modification required \\
\addlinespace
NL-to-PromQL & Metric discovery & Sub-second ($<$200ms catalog lookup) \\
(Sisodia) & Full pipeline & $\sim$1.1s end-to-end; $\sim$2,000 metric catalog \\
\bottomrule
\end{tabular}
\end{table}

\section{Synthesis: The State of AI Observability in 2026}

\subsection{What We Can Monitor Today}

The surveyed papers demonstrate that significant progress has been made at each individual layer:

\begin{itemize}[leftmargin=*]
\item \textbf{Model internals} can be probed for faithful world states even when outputs are adversarially corrupted (Berkeley).
\item \textbf{Confidence calibration} can be dramatically improved through modified training objectives without sacrificing accuracy (MIT).
\item \textbf{Behavioral reasoning} can be monitored at scale, with empirical scaling laws that guide deployment decisions (OpenAI).
\item \textbf{Operational agents} can be systematically evaluated with reproducible benchmarks (AIOpsLab).
\item \textbf{Infrastructure execution} can be traced end-to-end without code modification (TRUFFLD).
\item \textbf{Operational telemetry} can be synthesized into natural language insights through catalog-driven metric discovery and LLM summarization \citep{sisodia2026nlpromql}.
\end{itemize}

\subsection{What Remains Unsolved}

Four critical gaps define the frontier of AI observability research:

\paragraph{Gap 1: Cross-Layer Signal Correlation.} No existing system correlates a model's internal confidence (Layer 2) with infrastructure anomalies (Layer 5). If a model reports low confidence on a particular query, is that because the query is genuinely difficult, or because GPU memory pressure is causing KV cache evictions? Answering this question requires connecting signals that currently live in separate monitoring silos.

\paragraph{Gap 2: Unified Evaluation Benchmarks.} While AIOpsLab provides evaluation infrastructure for operational agents, there is no equivalent for model-level monitoring. Comparing RLCR's calibration improvements with propositional probe faithfulness requires a shared task definition, which does not yet exist.

\paragraph{Gap 3: Real-Time Adaptive Monitoring.} Current approaches are largely static: probes are trained offline, monitoring thresholds are set manually, and evaluation suites are run periodically. Production LLM systems need adaptive monitoring that adjusts its sensitivity based on current workload, model behavior, and environmental conditions.

\paragraph{Gap 4: Cost-Aware Monitoring Allocation.} OpenAI's ``monitorability tax'' finding implies that monitoring has a real compute cost. No framework exists for optimizing the allocation of monitoring resources across layers---deciding how much compute to spend on CoT analysis versus infrastructure tracing versus confidence calibration for a given deployment scenario.

\section{Future Directions}
\label{sec:future}

Based on our analysis, we propose four research directions that could advance the field toward integrated AI observability:

\paragraph{1. Vertical Integration Frameworks.} Building systems that consume signals from all five layers and correlate them in real time. Initial work on catalog-driven metric discovery \citep{sisodia2026nlpromql} demonstrates that operational telemetry can be synthesized into natural language insights. Extending this approach to incorporate model-level signals (confidence, CoT properties) alongside infrastructure telemetry would create a more complete operational picture.

\paragraph{2. Observability-Aware Training.} RLCR demonstrates that training objectives can be modified to improve monitorability without sacrificing performance. Future work should explore training objectives that simultaneously optimize for accuracy, calibration, and CoT monitorability---producing models that are designed from the ground up to be observable.

\paragraph{3. Agentic Observability Workflows.} Combining AIOpsLab's agent evaluation framework with TRUFFLD's infrastructure tracing and operational intelligence tools could enable fully autonomous diagnostic agents that not only detect anomalies but investigate their root causes across the entire stack, from model behavior to GPU kernel execution.

\paragraph{4. Federated Monitoring for Multi-Model Systems.} Modern LLM deployments increasingly involve multiple models---routers, rankers, generators, validators---each requiring distinct monitoring approaches. Research on federated monitoring architectures that track information flow and quality degradation across model boundaries is essential as these systems grow in complexity.

\section{Conclusion}

The AI observability landscape in 2026 is characterized by impressive depth at individual layers but limited integration across them. The five papers analyzed in this survey collectively demonstrate that LLM systems can be monitored at every level---from internal activations to GPU kernels---with methods ranging from interpretability probes to non-intrusive hardware tracing. The critical challenge ahead is building unified systems that connect these signals into coherent, actionable operational intelligence.

The practical deployment of such systems will require not only technical integration but also the development of new abstractions: shared data models for cross-layer telemetry, cost-aware monitoring allocation strategies, and evaluation frameworks that assess monitoring effectiveness end-to-end rather than layer-by-layer. The foundations are in place; the integration work is the next frontier.

\bibliography{references}

\end{document}